\documentstyle{l-aa}

\begin{document}

\newcommand{\equ}{\begin{equation}}
\newcommand{\eequ}{\end{equation}}
\newcommand{\arry}{\begin{eqnarray}}
\newcommand{\earry}{\end{eqnarray}}
\newcommand{\BF}{\begin{fig[Bure}}
\newcommand{\EF}{\end{figure}}
\newcommand{\BI}{\begin{itemize}}
\newcommand{\EI}{\end{itemize}}
\newcommand{\BE}{\begin{enumerate}}
\newcommand{\EE}{\end{enumerate}}
\newcommand{\dis}{\displaystyle}
\newcommand{\BC}{\begin{center}}
\newcommand{\EC}{\end{center}}
\newcommand{\BL}{\begin{flushleft}}
\newcommand{\EL}{\end{flushleft}}
\newcommand{\BTA}{\begin{table}}
\newcommand{\ETA}{\end{table}}
\newcommand{\BT}{\begin{tabbing}}
\newcommand{\ET}{\end{tabbing}}
\newcommand{\TAB}{\begin{tabular}}
\newcommand{\ETAB}{\end{tabular}}  
\newcommand{\BD}{\begin{description}}
\newcommand{\ED}{\end{description}} 

\newcommand{\ApJ}{{ApJ }}
\newcommand{\AsA}{{A\&A }}
\newcommand{\AsJ}{{AJ }}
\newcommand{\Mn}{{MNRAS }}
\newcommand{\Asp}{{\em Astrophys. Space Sci.}}
\newcommand{\ApJS}{{ApJS }}
\newcommand{\AsAS}{{A\&AS }}
\newcommand{\JQSRT}{{\em J. Quant. Spectros. Radiat. Transfer}}
\newcommand{\ARAS}{{\em Ann. Rev. Astr. Ap.}}
\newcommand{\Via}{{\em Vistas in Astronomy}}
\newcommand{\ApJL}{{\em Astrophys. J. Lett.}}
\newcommand{\Na}{{\em Nature}}
\newcommand{\AnA}{{\em Ann. d'Ap.}}
\newcommand{\PhR}{{\em Phys. Rev.}}
\newcommand{\Nse}{{\em Nucl. Sci. Engng.}}
\newcommand{\Msait}{{Mem. Soc. Astron. Ital. }}

\newcommand{\Spe}{{Spectral Evolution of Galaxies}}
\newcommand{\RT}{{\em Radiative Transfer}}
\newcommand{\Ntt}{{\em Neutron Transport Theory}}
\newcommand{\Ppim}{{\em Physical Processes in the Intestellar Medium}}
\newcommand{\Iaunu}{{\em IAU Symp.\ 1991}}
\newcommand{\Iauoq}{{\em IAU Symp.\ No.\ 108}}
\newcommand{\Rmcr}{{\em Recent developments of Magellanic Cloud Research}}
\newcommand{\Sn}{{\em Stellar Nucleosynthesis}}
\newcommand{\Iauon}{{\em IAU Symp.\ No.\ 135 1989, }}
\newcommand{\NIM}{{\em Nebulae and Interstellar Matter}}
\newcommand{\Iauno}{{\em IAU Symp.\ No.\  1990, The Galactic and Extragalactic}}
\newcommand{\Lssp}{{\em Light Scattering by Small Particles}}
\newcommand{\AGAS}{{\em Astrophysics of Gaseous Nebulae and Active Galactic
Nuclei}}
\newcommand{\PGAS}{{\em Physics of Thermal Gaseous Nebulae}}
\newcommand{\ND}{{\em Numerical Data and Functional Relationships in Science
and Technology}}
\newcommand{\nrsc}{{\em New Results on Standard Candles}}
\newcommand{\RMAA}{{\em Rev. Mex. A.A.}}
\newcommand{\GEO}{{\em Geochim. et Cosmochim. Acta}}
\newcommand{\Pasp}{{PASP }}

\thesaurus{1(11.19.5; 08.07.1; 11.19.4; 11.05.2)}

\title{Indicators of star formation: 4000 \AA $\,$ break and
Balmer lines}

\author{B.M.~Poggianti \inst{1,2}
\and G.~Barbaro
\inst{3} }

\offprints{B.M.~Poggianti (RGO)}

\institute{Institute of Astronomy and Royal Greenwich Observatory, 
Madingley Road, Cambridge, United Kingdom, bianca@ast.cam.ac.uk \and 
Kapteyn Instituut, PO Box 800, 9700 AV Groningen, The Netherlands
\and Dipartimento di Astronomia, vicolo dell'Osservatorio 5, 35122
Padova, Italy, barbaro@astrpd.pd.astro.it}

\date{Received; accepted}

\maketitle

\begin{abstract}
The behaviour of the 4000 \AA $\,$ break index
and of the equivalent width of the main
Balmer lines 
is investigated a) for a single star as a function of effective temperature,
gravity and metallicity and b) for a single stellar population 
as a function of age and metallicity. 
Consequences for the interpretation of integrated spectra are presented.


\end{abstract}

\section{Introduction}

In order to study the stellar populations of galaxies it is useful to test
the capability of some spectral features of revealing the presence of stars of
different ages. 

The quantities here considered are the break at 4000 \AA $\,$ $D_{4000}$
and the equivalent widths (EW) 
of the Balmer lines, in particular $\rm H \delta$. 
Unlike several other spectral indices,
they do not require high quality spectra and
are therefore suitable for studying galaxies at intermediate and 
high redshifts.  

$D_{4000}$ is largely used to determine the
star formation characteristics of distant field and cluster galaxies. 
Hereafter the 4000 \AA $\,$ break is defined as
the ratio between the average flux density
in $\rm ergs \, s^{-1} cm^{-2} Hz^{-1}$ between 4050 and 4250 \AA $\,$
and that between 3750 and  3950 \AA $\,$ 
(Bruzual 1983).

The analysis of the Balmer lines has been employed in a countless
number of studies of stellar clusters, local and distant galaxies;
indices including Balmer lines are useful for
estimating the ages of star clusters, in the comparison
between globular clusters and ellipticals and in the
detection of recent starbursts in galaxies.
A strong absorption $\rm H\delta$ line has been detected in several cluster
galaxies with a wide range of colours
at $0 \le z <0.6$ (Couch \& Sharples 1987, Dressler \& Gunn 1992,
Caldwell et al. 1993, Belloni 
et al. 1995).
This feature  is usually interpreted as evidence of a burst of star formation
ended 0.5-1.5 
Gyr prior to the moment of observation (Dressler \& Gunn 1983), 
assuming implicitely that
the A-type stars responsible for the Balmer lines
are \em in the main sequence phase. \rm 
Here we investigate if stars in other evolutionary phases could give rise to
a strong $\rm H \delta$ line in order to determine if and when
this feature is an unambiguous sign of recent star formation.

Although largely used, these features lack a
systematic analysis (such a study exists for the $\rm H\beta$ indices, see
Buzzoni et al. 1994 and references therein), 
needed to clarify their dependence 
on metallicity (compare for instance Dressler \& Shectman (1987)--
Kimble et al. (1989) for the $D_{4000}$ index 
and Bica \& Alloin (1986)--Gregg(1994) for the Balmer lines).
This problem is related to the well known age-metallicity degeneracy, 
investigated by many authors especially for old stellar populations
(Jones \& Worthey 1995 and references therein). 
Compared to previous analysis, 
the spectral quantities here considered are influenced also by more recent
and even present star formation.

Considering stellar models instead of observed stellar libraries
allows one to investigate the whole
range of stellar parameters.
Single stellar population (SSP) and galaxy integrated 
spectra are computed  by means of an evolutionary synthesis model
that includes both the stellar 
contribution and the emission of the ionized gas (Barbaro \& Poggianti 1997). 
The advantages in taking into account a wide range of metallicities 
and the main advanced stellar stages,
in particular the horizontal branch (HB) phase,
will emerge clear from the subsequent discussion.
The problem of the "second parameter", i.e. the fact that stellar clusters
of comparable metallicities may have distinct horizontal branch
morphologies, is not taken into account, being this question not directly 
relevant to the present study (see de Freitas Pacheco \& Barbuy 1995
for an analysis of the $\rm H\beta$ indices of clusters of different HB 
morphologies).
The total (stellar absorption + gaseous emission) EWs
of the Balmer lines were measured by direct numerical 
integration, using the SPLOT program in IRAF and setting interactively
the continuum levels and the integration limits.
For SSPs and galaxies a Salpeter (1955)
Initial Mass Function (IMF) between 0.1 and 100 $M_{\odot}$ was assumed. 

\section{$D_{4000}$}

The variation of $D_{4000}$ with stellar spectral type and luminosity class
has been investigated by Hamilton (1985).
We have compared his results with  Kurucz's (1979)
models of solar metallicity using the spectral type-effective temperature and
luminosity class-surface gravity relations derived by  Schmidt-Kaler
(Landolt-B\"ornstein 1982, Tables 4.1.4.3 and 4.1.5.23). The comparison shows
that the theoretical values fit nicely the observed ones.
Using a later version of Kurucz's models does not alter the results
here presented.

The behaviour of $D_{4000}$ as a function of the
stellar parameters
(Fig. 1) has been derived from Kurucz's models.
For stars with $3500 <
T_{\rm eff} < 5500 \, \rm K$ (not included in Fig. 1) $D_{4000}$ has been
computed from the stellar spectra of Straizys \& Sviderskiene (1972) and
keeps always values above 2, reaching a maximum of 3 at $T_{\rm eff} 
\simeq 4000$ K.

\begin{figure}
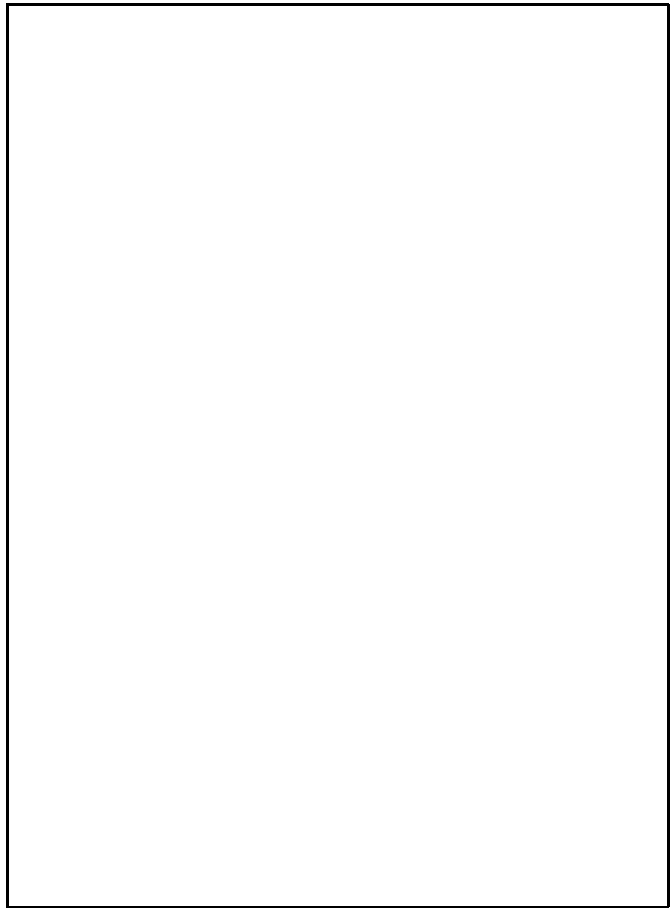

\picplace{12 cm}
\caption[]{$D_{4000}$ indices of Kurucz's 
stellar models: a) as a function of
the stellar effective temperature for solar metallicity and
two different gravities 
(filled circles and solid line Log g=4; empty circles 
and dotted line Log g=1.5); b) as a function of the gravity 
for solar metallicity stars and 4 different effective temperatures;
c) as a function of the metallicity for Log g =4 stars and two different
effective temperatures}
\end{figure}

The $D_{4000}$ values for SSPs, computed using Kurucz's spectra,
are shown in Fig. 2; the typical uncertainty is  0.1.
While this index raises to high values for old solar metallicity
SSPs, its value remains much lower in the case of low metallicity SSPs
even after 15 Gyr.

The results obtained from the integrated spectra of elliptical models 
with different average metallicities are shown in
Table 1, while those of other Hubble types are given in Table 2.
It must be stressed that the results presented in the tables
include \em both \rm metallicity and age effects:
in computing the integrated spectrum, stellar populations of different 
metallicities are taken into account, according to a standard
chemical evolutionary model that uses the instantaneous recycling approximation.
A full description of the adopted galactic models is given in Barbaro \&
Poggianti (1997). 

\begin{figure}
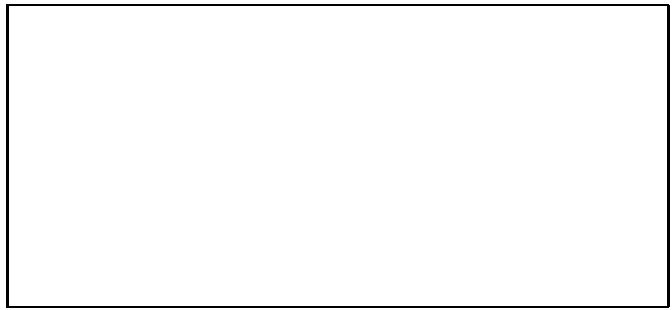

\picplace{4 cm}
\caption[]{$D_{4000}$ values of single stellar populations}
\end{figure}

\BTA

\caption[]{$D_{4000}$ of the elliptical model as a function of the average 
metallicity}

\begin{flushleft}

\TAB{rr}          
\hline 
\noalign{\smallskip}
Z & $D_{4000}$ \\
\noalign{\smallskip}
\hline
\noalign{\smallskip}
0.03  & 2.26 \\
0.02  & 2.21 \\
0.01  & 2.07 \\
0.008 & 2.03 \\
0.005 & 1.94 \\
0.003 & 1.85 \\
\noalign{\smallskip}                                                         
\hline
\ETAB

\end{flushleft}

\ETA

\BTA

\caption[]{Results for $D_{4000}$ and $\rm EW(H\delta) $ including
the gaseous emission (total) and only for the stellar component (only abs.)}

\begin{flushleft}

\TAB{llll}          
\hline 
\noalign{\smallskip}
Type & $D_{4000}$ & $\rm EW(H\delta) $ & $\rm EW(H\delta)$ \\
\noalign{\smallskip}
\hline
\noalign{\smallskip}
     &            &  total & only abs. \\
\noalign{\smallskip}
\hline
\noalign{\smallskip}
E & 2.21 & 0.9 & 0.9 \\
Sa & 1.91 & 1.1 & 1.2 \\
Sb & 1.75 & 1.0 & 2.0 \\
Sc & 1.54 & 1.4 & 3.2 \\
Sd & 1.42 & 1.7 & 3.7 \\
Ex & 1.31 & 1.2 & 4.1 \\
\noalign{\smallskip}
\hline
\ETAB

\end{flushleft}

\ETA

\section{$\rm H\delta$ and other Balmer lines}

Figure 3 illustrates the behaviour of EW($\rm H\delta$) as a function of the 
stellar parameters; values are taken from Table 8 by Kurucz (1979).
The $\rm H\delta$ line is strong in a range of temperatures (Fig. 3a), 
corresponding approximately to a lifetime interval of
 0.2-2 Gyr on the main sequence. Moreover Fig. 3 shows that 
high EWs are reached in
that range of temperatures only for high surface gravities (Fig. 3b).
The maximum $\rm H\delta$ is therefore reached for an A0V-type star.
\begin{figure}
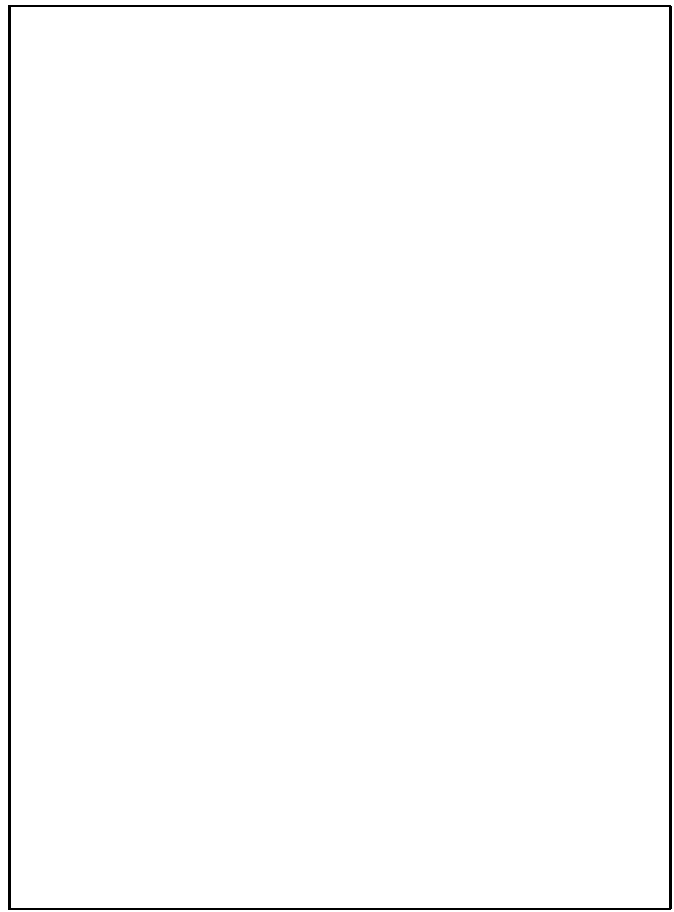

\picplace{12 cm}
\caption[]{$\rm EW(H\delta)$ of single stars: a)
as a function of $T_{\rm eff}$ for solar metallicity and two different
gravities (filled circles and solid line Log g=4;
empty circles and dotted line Log g=1.5);
b) as a function of the gravity for solar metallicity and 4 different 
effective temperatures; c) as a function of the metallicity for 2 different
gravities and 2 different effective temperatures}
\end{figure}
The $\rm H\alpha, \, H\beta \, and \, H\gamma$ lines show the same
behaviour of $\rm H\delta$, the only difference being a decrease of the maximum
EW at lower order lines;
all the following conclusions are therefore
valid also for the other Balmer lines.

\begin{figure}
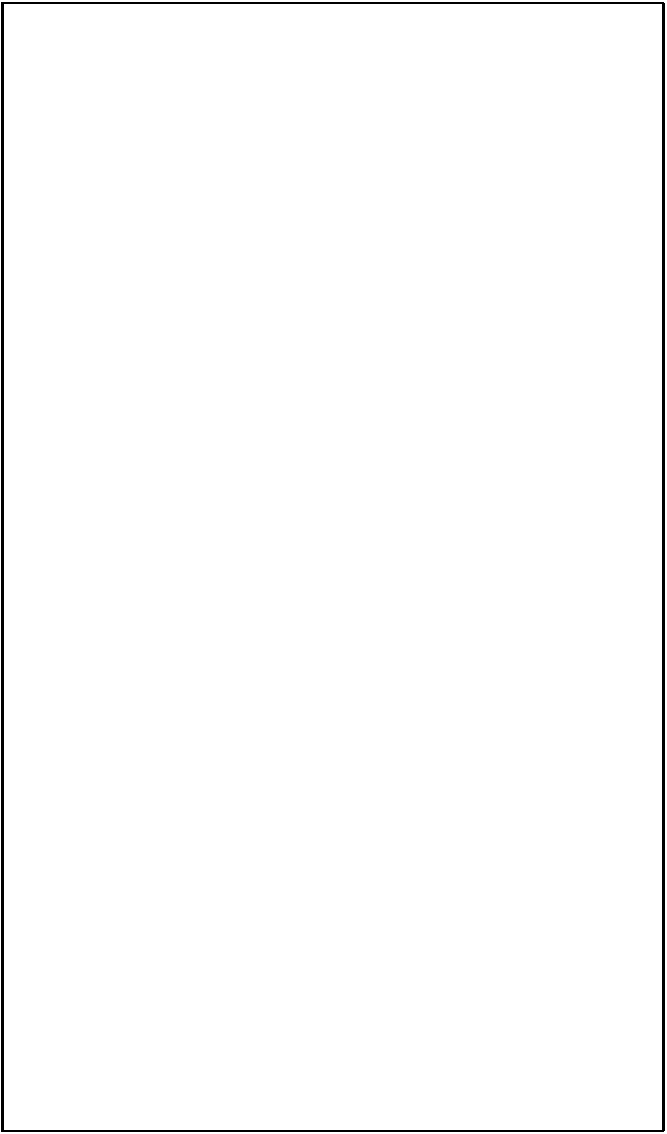

\picplace{15 cm}
\caption[]{14 Gyr old isochrones of Z=0.02 (top), Z=0.001 (center) and
Z=0.0001 (bottom). Rectangles show the region of highest EW($\rm H\delta$);
the highest values are reached in the inner rectangle.
The zero age $Z=Z_{\odot}$ main sequence is also shown (empty circles)}
\end{figure}

In order to verify if only main sequence stars have the
range of temperatures and the high gravities necessary to originate strong
Balmer lines, old isochrones (14 Gyr) of different metallicities are
shown in Fig. 4 on a modified HR diagram (log g versus $T_{\rm eff})$.
These diagrams reproduce only the isochrones and do not give
the density distribution along them according to the IMF.
The main evolutionary phases are easily recognized (main sequence, 
giant branch, HB, AGB and Post-AGB). 
The hottest Post-AGB stars lie
out of the diagram on the left-bottom side (high $T_{\rm eff}$ and g)
and therefore cannot contribute significantly to the $\rm H\delta$ line.
The only stars able to reach the $\rm H \delta$-strong region of 
Fig. 4 are the bluest horizontal branch stars with low Z.
Stars with solar metallicity do not reach high enough temperatures in this phase.
On the contrary, for an age 800 Myr
the stellar objects in the rectangles are
those around the main sequence turn-off  ($1.5-2 M_{\odot}$),
regardless of Z.

In extremely metal-rich populations, other types of stars are expected to
have the required temperatures and gravities: Hot and Very Hot HB 
objects and AGB-manqu\`e stars (Greggio \& Renzini 1990, Liebert et al. 1994, 
Whitney et al. 1994, Bressan et al. 1994).
Their contribution to the integrated spectra of stellar systems needs
to be verified. On the other hand, if metal-rich HB
stars would be responsible for the observed strong Balmer
lines in some galaxies, these should belong to
a restricted class of objects (the most metal-rich and most
luminous) and this luminosity selection is not observed.

\begin{figure}
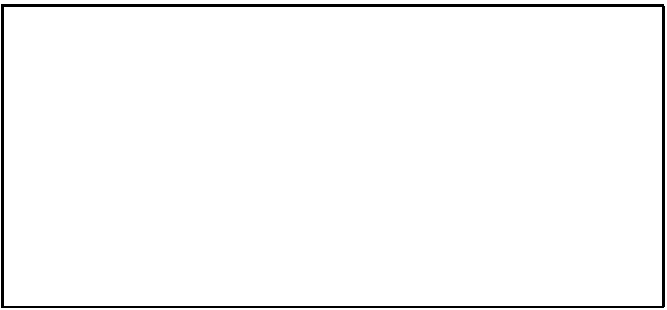

\picplace{4 cm}
\caption[]{$\rm EW(H\delta)$ of single stellar populations of solar 
metallicity as a function of the age of the SSP; in this case the gaseous
emission is not taken into account}
\end{figure}

The $\rm EW(H \delta)$ of $\rm Z=Z_{\odot}$ SSPs of different
ages are shown in Fig. 5; they are computed using the Jacoby et al. 
stellar library (1984), whose higher spectral resolution (4 \AA)
compared to Kurucz's models (20 \AA) allows one to study
spectral features such as the Balmer lines.
The typical uncertainty on the measure of EW is 0.2 \AA.
If the gaseous emission is not considered, $\rm EW(H\delta)$
reaches values  $> 3$ \AA $\,$ already at 3 Myr and remains strong
for about 1.5 Gyr;
the maximum is reached at about 300 Myr. Comparing our results
with the Balmer lines observed in globular clusters
(Bica \& Alloin 1986, Cohen et al. 1984), a very good agreement
is found.
In older SSPs (5-15 Gyr), $\rm EW(H\delta)=0.5 \pm 0.1$ \AA.
If the gaseous contribution is included, the line is in emission during the 
first 10 Myr (having a value of about 40 \AA $\,$ during the first 3 Myr), 
while the values at the following times remains unchanged.


The stars of the spectral library employed have $Z \sim
Z_{\odot}$, therefore in principle the metallicity dependence of the
results could not be investigated. We have seen however
(Fig. 3c) that, given the 
effective temperature and surface gravity, the $\rm EW(H \delta)$
does not depend on the metallicity of the atmosphere.
The internal structure, and consequently the position of the stars
of an SSP on the HR diagram, are instead
strongly dependent on the metal content.
Therefore the
strong Balmer lines observed in the integrated spectra of globular clusters 
can be studied also using the stellar library of
Jacoby et al., by considering isochrones of different metallicities and 
adopting the Jacoby et al. spectra for any isochrone metal content. 
The main difference with 
respect to the solar metallicity case is that the $\rm EW(H \delta)$
at low Z (0.0001)
is not negligible also for old populations,
due to the presence of hot HB stars previously discussed.
The dependence of the EW of two Balmer lines on the
metallicity for an old SSP (14 Gyr) is shown in Fig. 6.
For the $\rm H\delta$ line two distinct regimes are clearly visible
and the relations are linear with a correlation coefficient greater than
98 \%:
\equ
\rm
EW(H\delta) (\AA)= -1115.8 \cdot Z + 3.1  \, \, \, if \, 0.0001 \le Z \le 0.001
\eequ
\equ
\rm
EW(H\delta) (\AA)= -21.0 \cdot Z + 2.0 \, \, \, \,    if \, 
0.003 \le Z \le 0.03
\eequ


\begin{figure}
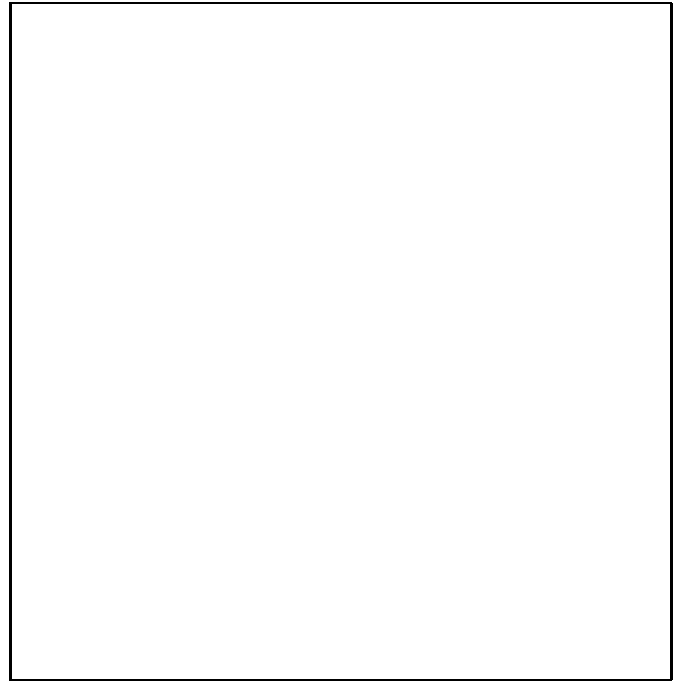

\picplace{9 cm}
\caption[]{EW(H$\delta$) and EW(H$\beta$) of 14 Gyr old SSPs as a function of
the metallicity. [Fe/H] values are computed from Z using solar 
abundances ratios.}
\end{figure}
The strong gradient at low metallicities and the shallower at 
$Z>0.003$ are the direct consequence of the non-monotonic
behaviour of HB 
stars with metallicity illustrated in Fig. 4.
This can account for the large spread in $\rm EW(H\delta)$
observed by Bica \& Alloin (1986) for old globular clusters.

In the case of the $H\beta$ line (Fig. 6c,d), after the steep gradient
at low Z, the line is independent of metallicity for $Z>0.005$.
This could account for the large scatter 
and the lack of any 
correlation of the Balmer lines with metallicity above [Fe/H]=-1 reported by 
Gregg (1993), Faber \& Worthey (1993) and Gregg (1994), for which alternative
hypotheses have been proposed but seem unsatisfactory
(interstellar $H\beta$ emission filling,
blue stragglers in post-core-collapse clusters).

\section{Conclusions}

This study confirms that
spectral indices or equivalent widhts of $\rm H \beta$ and of the other Balmer
lines depend on metallicity in SSPs and galaxies and therefore cannot be used
as pure age indicator. On the other side, this metallicity dependence 
can account for variations of the equivalent widths up to certain values
(typically 1.5 \AA $\,$ for H$\delta$ and H$\beta$ for old SSPs) and 
above such values age differences need to be taken into account.

Results for galaxy models are shown in Table 2. In spirals
the absorption line is partially filled in by the gaseous emission;
the separate contributions are given as well.
The very metal poor populations are unable to give rise to
strong  $\rm EW(H\delta)$ in the integrated spectrum of galaxies.
This confirms that an $\rm EW(H\delta)$ as high as those observed 
in distant cluster galaxies is the signature
of an enhanced star formation episode in the last 2 Gyr and cannot be due
to metallicity.

\begin{acknowledgements}
We acknowledge the availability of the Jacoby et al's stellar
library from the NDSS-DCA Astronomical Data Center.
This work was supported in part by the Formation and Evolution of
Galaxies network set up by the European Commission under contract
ERB FMRX-CT96-086 of its TMR programme.
B.M.P. acknowledges the receipt of a grant from the Department of Physics
of the University of Pisa.
\end{acknowledgements}

%


\begin{thebibliography}{}

\bibitem{}
Barbaro G., Poggianti B.M., 1997, \AsA in press
\bibitem{}
Belloni P., Bruzual G.A., R\"oser H.J., Thimm G.J., 1995, \AsA 297, 61
\bibitem{}
Bica E., Alloin D., 1986, \AsA 162, 21
\bibitem{}
Bressan A.G., Chiosi C., Fagotto F., 1994, \ApJS 94, 63
\bibitem{}
Bruzual A.G., 1983, \ApJ 273, 105 
\bibitem{}
Buzzoni A., Mantegazza L., Gariboldi G., 1994, \AsJ 107, 513
\bibitem{}
Caldwell N., Rose J.A., Sharples R.M., Ellis R.S., Bower R.G., 1993, \AsJ
106, 473
\bibitem{}
Cohen J.G., Persson S.E., Searle L., 1984, \ApJ 281, 141 
\bibitem{}
Couch W.J., Sharples R.M., 1987, \Mn 229, 423
\bibitem{}
de Freitas Pacheco J.A., Barbuy B., 1995, \AsA 302, 718
\bibitem{}
Dressler A., Gunn J.E., 1983, \ApJ 270, 7
\bibitem{}
Dressler A., Gunn J.E., 1992, \ApJS 78, 1 
\bibitem{}
Dressler A., Shectman S.A., 1987, \AsJ 94, 899  
\bibitem{}
Faber S.M., Worthey G., 1993, Spectral Features in Globular Clusters and
Elliptical Galaxies. In: Smith G.H., Brodie J.P. (eds.) The Globular Cluster-Galaxy
Connection. ASP Conf. Ser. 48, 508
\bibitem{}
Gregg M.D., 1994, \AsJ 108, 2164
\bibitem{}
Gregg M.D., 1993, Comparison of Elliptical Galaxy and Metal-Rich Globular 
Cluster Spectra. In: Smith G.H., Brodie J.P. (eds.) The Globular Cluster-Galaxy
Connection. ASP Conf. Ser. 48, 565
\bibitem{}
Greggio L., Renzini A., 1990, \ApJ 364, 35
\bibitem{}
Hamilton D., 1985, \ApJ 297, 371 
\bibitem{}
Jacoby G.H., Hunter D.A., Christian C.A., 1984, \ApJS 56, 257 
\bibitem{}
Jones L.A., Worthey G., 1995, \ApJ 446, L31
\bibitem{}
Kimble R.A., Davidsen A.F., Sandage A.R., 1989, Ap\&SS 157, 237
\bibitem{}
Kurucz R., 1979, \ApJS 40, 1 
\bibitem{}
Landolt--B\"ornstein: 1982, {\it Numerical Data and Functional
Relationships in Science and Technology}, Springer-Verlag, vol.2, 
subvol. b 
\bibitem{}
Liebert J., Saffer R.A., Green E.M., 1994, \AsJ  107, 1408 
\bibitem{}
Salpeter E.E., 1955, \ApJ 121, 161
\bibitem{}
Straizys V., Sviderskiene Z., 1972, {\it Bull. Vilnius An. Obs.} 35, 1 
\bibitem{}
Whitney J.H., O'Connell R.W., Rood R.T., 1994 \AsJ 108, 1350

\end{thebibliography}
\end{document}